\documentclass[twocolumn,
              superscriptaddress,
              prl, aps,
              showkeys,
              showpacs,
              nofootinbib,
              notitlepage,
              floatfix,
              longbibliography,
              ]{revtex4-2}

\usepackage[english]{babel}
\usepackage[letterpaper,top=2cm,bottom=2cm,left=2cm,right=2cm,marginparwidth=1.75cm]{geometry}
\usepackage{amsmath}
\usepackage{graphicx}
\usepackage{booktabs}
\usepackage{siunitx}
\usepackage{upgreek}
\usepackage{xspace}
\usepackage[colorlinks=true,citecolor=blue,filecolor=blue,linkcolor=blue,urlcolor=blue,pdftex]{hyperref}
\usepackage[dvipsnames]{xcolor}
\usepackage{lipsum}
\usepackage{multirow}
\usepackage{hyperref}
\usepackage[capitalise]{cleveref}
\crefname{section}{Sec.}{Secs.}
\crefname{table}{Tab.}{Tabs.} 
\usepackage{subfigure}
\usepackage{adjustbox}
\usepackage{textcomp}
\usepackage{amsmath,amsfonts,amsthm,bm}
\usepackage{orcidlink} 
\usepackage{currfile} 

\usepackage{lineno}
\DeclareSIUnit\eVperc{\eV\per\clight}
\DeclareSIUnit\clight{\text{\ensuremath{c}}}

\begin{document}



\newcommand{\bologna}{\affiliation{Department of Physics and Astronomy, University of Bologna and INFN-Bologna, 40126 Bologna, Italy}}
\newcommand{\chicago}{\affiliation{Department of Physics \& Kavli Institute for Cosmological Physics, University of Chicago, Chicago, IL 60637, USA}}
\newcommand{\coimbra}{\affiliation{LIBPhys, Department of Physics, University of Coimbra, 3004-516 Coimbra, Portugal}}
\newcommand{\columbia}{\affiliation{Physics Department, Columbia University, New York, NY 10027, USA}}
\newcommand{\lngs}{\affiliation{INFN-Laboratori Nazionali del Gran Sasso and Gran Sasso Science Institute, 67100 L'Aquila, Italy}}
\newcommand{\mainz}{\affiliation{Institut f\"ur Physik \& Exzellenzcluster PRISMA$^{+}$, Johannes Gutenberg-Universit\"at Mainz, 55099 Mainz, Germany}}
\newcommand{\mpik}{\affiliation{Max-Planck-Institut f\"ur Kernphysik, 69117 Heidelberg, Germany}}
\newcommand{\munster}{\affiliation{Institut f\"ur Kernphysik, Westf\"alische Wilhelms-Universit\"at M\"unster, 48149 M\"unster, Germany}}
\newcommand{\nikhef}{\affiliation{Nikhef and the University of Amsterdam, Science Park, 1098XG Amsterdam, Netherlands}}
\newcommand{\nyuad}{\affiliation{New York University Abu Dhabi - Center for Astro, Particle and Planetary Physics, Abu Dhabi, United Arab Emirates}}
\newcommand{\purdue}{\affiliation{Department of Physics and Astronomy, Purdue University, West Lafayette, IN 47907, USA}}
\newcommand{\rice}{\affiliation{Department of Physics and Astronomy, Rice University, Houston, TX 77005, USA}}
\newcommand{\stockholm}{\affiliation{Oskar Klein Centre, Department of Physics, Stockholm University, AlbaNova, Stockholm SE-10691, Sweden}}
\newcommand{\subatech}{\affiliation{SUBATECH, IMT Atlantique, CNRS/IN2P3, Universit\'e de Nantes, Nantes 44307, France}}
\newcommand{\torino}{\affiliation{INAF-Astrophysical Observatory of Torino, Department of Physics, University  of  Torino and  INFN-Torino,  10125  Torino,  Italy}}
\newcommand{\ucsd}{\affiliation{Department of Physics, University of California San Diego, La Jolla, CA 92093, USA}}
\newcommand{\wis}{\affiliation{Department of Particle Physics and Astrophysics, Weizmann Institute of Science, Rehovot 7610001, Israel}}
\newcommand{\zurich}{\affiliation{Physik-Institut, University of Z\"urich, 8057  Z\"urich, Switzerland}}
\newcommand{\paris}{\affiliation{LPNHE, Sorbonne Universit\'{e}, CNRS/IN2P3, 75005 Paris, France}}
\newcommand{\freiburg}{\affiliation{Physikalisches Institut, Universit\"at Freiburg, 79104 Freiburg, Germany}}
\newcommand{\napels}{\affiliation{Department of Physics ``Ettore Pancini'', University of Napoli and INFN-Napoli, 80126 Napoli, Italy}}
\newcommand{\nagoya}{\affiliation{Kobayashi-Maskawa Institute for the Origin of Particles and the Universe, and Institute for Space-Earth Environmental Research, Nagoya University, Furo-cho, Chikusa-ku, Nagoya, Aichi 464-8602, Japan}}
\newcommand{\laquila}{\affiliation{Department of Physics and Chemistry, University of L'Aquila, 67100 L'Aquila, Italy}}
\newcommand{\tokyo}{\affiliation{Kamioka Observatory, Institute for Cosmic Ray Research, and Kavli Institute for the Physics and Mathematics of the Universe (WPI), University of Tokyo, Higashi-Mozumi, Kamioka, Hida, Gifu 506-1205, Japan}}
\newcommand{\kobe}{\affiliation{Department of Physics, Kobe University, Kobe, Hyogo 657-8501, Japan}}
\newcommand{\kit}{\affiliation{Institute for Astroparticle Physics, Karlsruhe Institute of Technology, 76021 Karlsruhe, Germany}}
\newcommand{\tsinghua}{\affiliation{Department of Physics \& Center for High Energy Physics, Tsinghua University, Beijing 100084, P.R. China}}
\newcommand{\ferrara}{\affiliation{INFN-Ferrara and Dip. di Fisica e Scienze della Terra, Universit\`a di Ferrara, 44122 Ferrara, Italy}}
\newcommand{\groningen}{\affiliation{Nikhef and the University of Groningen, Van Swinderen Institute, 9747AG Groningen, Netherlands}}
\newcommand{\westlake}{\affiliation{Department of Physics, School of Science, Westlake University, Hangzhou 310030, P.R. China}}
\newcommand{\shenzhen}{\affiliation{School of Science and Engineering, The Chinese University of Hong Kong, Shenzhen, Guangdong, 518172, P.R. China}}
\newcommand{\coimbrapoli}{\affiliation{Coimbra Polytechnic - ISEC, 3030-199 Coimbra, Portugal}}
\newcommand{\heidelberg}{\affiliation{Physikalisches Institut, Universit\"at Heidelberg, Heidelberg, Germany}}
\newcommand{\roma}{\affiliation{INFN-Roma Tre, 00146 Roma, Italy}}
\newcommand{\bucknell}{\affiliation{Department of Physics \& Astronomy, Bucknell University, Lewisburg, PA, USA}}
\newcommand{\isct}{\affiliation{Department of Physics, School of Science, Institute of Science Tokyo, Meguro, Tokyo, 152-8551, Japan}}


\author{E.~Aprile\,\orcidlink{0000-0001-6595-7098}}\columbia
\author{J.~Aalbers\,\orcidlink{0000-0003-0030-0030}}\groningen
\author{K.~Abe\,\orcidlink{0009-0000-9620-788X}}\tokyo
\author{M.~Abu~Rmilah\,\orcidlink{0009-0007-9750-6655}}\wis
\author{M.~Adrover\,\orcidlink{0123-4567-8901-2345}}\zurich
\author{S.~Ahmed~Maouloud\,\orcidlink{0000-0002-0844-4576}}\paris
\author{L.~Althueser\,\orcidlink{0000-0002-5468-4298}}\munster
\author{B.~Andrieu\,\orcidlink{0009-0002-6485-4163}}\paris
\author{E.~Angelino\,\orcidlink{0000-0002-6695-4355}}\lngs\chicago
\author{D.~Ant\'on~Martin\,\orcidlink{0000-0001-7725-5552}}\email[]{dantonmartin@uchicago.edu}\chicago
% ...
\collaboration{XENON Collaboration}\email[]{xenon@lngs.infn.it}\noaffiliation

%

\title{Search for Magnetic and Spin-Independent Inelastic Dark Matter with XENONnT}

\newcommand*{\comment}{\textcolor{red}}
\newcommand*{\needs}{\textcolor{red}}
\newcommand{\rntwotwozero}{\ensuremath{^{220}\mathrm{Rn}}\xspace}
\newcommand{\toymc}{toy-MC\xspace}
\newcommand{\toymcs}{toy-MCs\xspace}
\newcommand{\cevns}{CE\ensuremath{\nu}NS\xspace}
\newcommand{\x}{\ensuremath{\mathrm{X}}\xspace}
\newcommand{\y}{\ensuremath{\mathrm{Y}}\xspace}
\newcommand{\z}{\ensuremath{\mathrm{Z}}\xspace}
\newcommand{\R}{\ensuremath{\mathrm{R}}\xspace}
\newcommand{\radius}{\ensuremath{\mathrm{R}}\xspace}
\newcommand{\wimp}{\ensuremath{\mathrm{WIMP}}\xspace}

\newcommand{\expectationACnonwirenominaloffall}{ $2.0 \pm 0.6 $ }
\newcommand{\expectationACwirenominalonall}{ $2.3 \pm 0.7 $ }
\newcommand{\expectationatnunominalbothall}{ $0.05 \pm 0.02 $ }
\newcommand{\expectationcevnsnominalbothall}{ $0.19 \pm 0.06 $ }
\newcommand{\expectationernominalbothall}{ $134.5 $ }
\newcommand{\expectationradiogenicnominalbothall}{ $0.8 \pm 0.4 $ }
\newcommand{\expectationradiogenicXnominalbothall}{ $0.31 \pm 0.16 $ }
\newcommand{\expectationwallnominalbothall}{ $14 \pm 3 $ }
\newcommand{\expectationercalibrationnominalboth}{ $2062 \pm 210 $ }
\newcommand{\expectationparonenominal}{ $0.0 $ }
\newcommand{\expectationpartwonominal}{ $0.0 $ }
\newcommand{\expectationACnonwirebestoffall}{ $2.1 \pm 0.6 $ }
\newcommand{\expectationACwirebestonall}{ $2.3 \pm 0.6 $ }
\newcommand{\expectationatnubestbothall}{ $0.04 \pm 0.02 $ }
\newcommand{\expectationcevnsbestbothall}{ $0.19 \pm 0.06 $ }
\newcommand{\expectationerbestbothall}{ $135_{ -11  }^{ +12 }$ }
\newcommand{\expectationradiogenicbestbothall}{ $0.8 \pm 0.4 $ }
\newcommand{\expectationradiogenicXbestbothall}{ $0.30 \pm 0.15 $ }
\newcommand{\expectationwallbestbothall}{ $12 \pm 2 $ }
\newcommand{\expectationercalibrationbestboth}{ $2052 \pm 44 $ }
\newcommand{\expectationparonebest}{ $0.4 \pm 0.2 $ }
\newcommand{\expectationpartwobest}{ $-1.8_{ -0.6  }^{ +0.7 }$ }

\makeatletter
\newcommand{\fmarki}{*}
\newcommand{\fmarkii}{\ensuremath{\dagger}}
\newcommand{\fmarkiii}{\ensuremath{\ddagger}}
\newcommand{\fmarkiv}{\ensuremath{\mathsection}}
\newcommand{\fmarkv}{\ensuremath{\mathparagraph}}
\newcommand{\fmarkvi}{\ensuremath{\|}}
\newcommand{\fmarkvii}{**}
\newcommand{\fmarkviii}{\ensuremath{\dagger\dagger}}
\newcommand{\fmarkix}{\ensuremath{\ddagger\ddagger}}

\def\@fnsymbol#1{{\ifcase#1\or \fmarki\or \fmarkii\or \fmarkiii\or \fmarkiv\or \fmarkv\or \fmarkvi\or \fmarkvii\or \fmarkviii\or \fmarkix \else\@ctrerr\fi}}
\makeatother
\date{\today}

\begin{abstract}
    We present a search for Magnetic and Spin-Independent inelastic Dark Matter  using 2.1 tonne-years of data from the XENONnT experiment. 
    We consider both single- and double-site event topologies, targeting the unique signature of an initial nuclear recoil followed by a delayed de-excitation photon ($\chi^* \rightarrow \chi + \gamma$). 
    To suppress backgrounds, we introduce novel directional and kinematic selections based on the inferred speed and direction of $\chi^*$ between the scatter and decay sites. 
    We find that the collected data are consistent with background expectations, and report 90\% C.L. upper limits for both models across the GeV/c$^2$--TeV/c$^2$ mass range.
    
\end{abstract}

\keywords{inelastic Dark Matter, Direct Detection, Xenon}

\maketitle

\textit{Introduction}---Indirect observational evidence from astrophysics and cosmology suggests that approximately 85\% of the matter content of the Universe consists of Dark Matter (DM), the fundamental nature of which remains unresolved \cite{Planck2020}. Among direct Dark Matter detection techniques, Liquid Xenon Time Projection Chambers (LXeTPC) ~\cite{XENONnT:2025sr1, LZ:2022ufs, PandaX-4T:2021bab} have established the most stringent constraints on DM candidates in the GeV/c$^2$--TeV/c$^2$ mass range, with the Weakly Interacting Massive Particle (WIMP) representing the prevailing theoretical paradigm~\cite{Bertone:2004pz}. In its simplest formulation, the WIMP hypothesis assumes a single DM particle, such that elastic scattering off nuclei constitutes the primary observable signature. In this Letter, we relax this assumption and leverage the target mass and ultra-low background of XENONnT to investigate two classes of inelastic DM models: Magnetic inelastic DM (MiDM) \cite{Chang:2010en} and Spin-Independent inelastic DM (SIiDM) \cite{TuckerSmith:2001hy}. Models of inelastic DM are of interest due to their characteristic recoil spectra and signatures, as well as due to the well-motivated theory behind them \cite{PhysRevD.86.075021, garcia2025minimalisticmodelinelasticdark}. 

\begin{figure}[!h]
    \centering
    \includegraphics[width=0.65\linewidth]{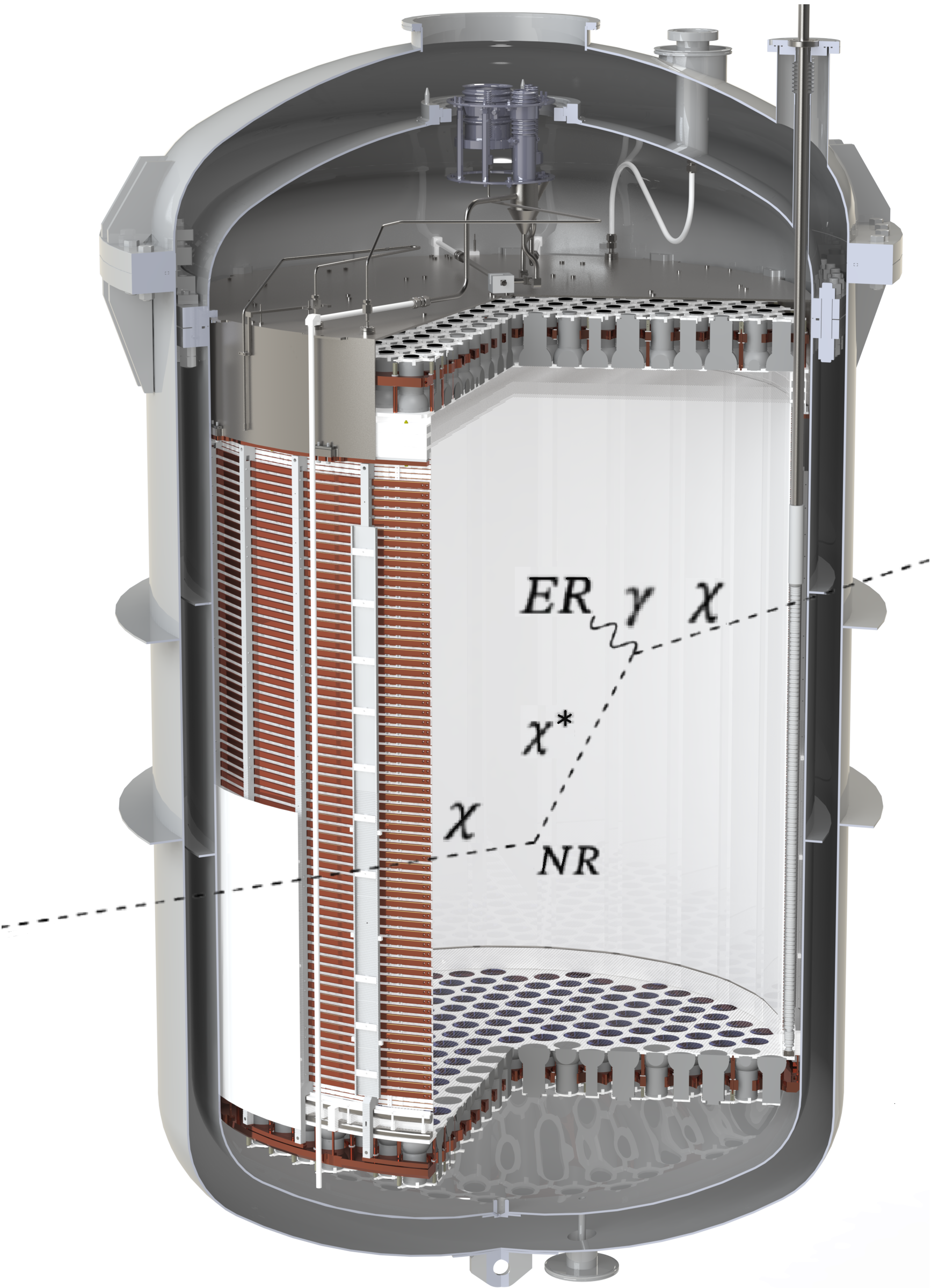}   
    \raisebox{28mm}{    \includegraphics[width=0.65\linewidth,angle=270,origin=c]{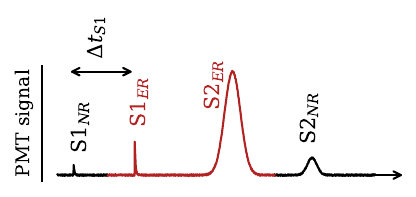}
    }
    \caption{Exemplary double-site MiDM signal in the XENONnT detector (left); the resulting waveform topology is depicted on the right. The ER S1 and S2 are larger than their NR counterparts, the NR S1 always precedes the ER S1, and the time ordering of the S2s depends on the z-coordinate of the interactions.}
    \label{fig:topology}
\end{figure}

MiDM and SIiDM are characterized by the existence of two mass states of DM, denoted as $\chi$ and $\chi^*$,  with a mass difference between the two of size $\delta \equiv m_{\chi^*} - m_{\chi}$. The interaction with ordinary matter is parameterized by a magnetic dipole moment $\mu_\chi$ for MiDM, 
and by a spin-independent nucleon cross section $\sigma_{\mathrm{SI}}$ for SIiDM. MiDM couples to both the magnetic moment (dipole-dipole, DD) and the electric charge (dipole-charge, DZ) of the target xenon nuclei. Interactions are mediated by off-diagonal terms, such that a transition $\chi \rightarrow \chi^*$ occurs during the scattering. This endothermic process imposes a strict kinematic threshold: to deposit a nuclear recoil energy $E_{\mathrm{R}}$, the incident DM velocity must exceed a minimum value $v_{\min}$:
\begin{equation}
        v_{\min} = \frac{1}{\sqrt{2 m_{\mathrm{N}} E_{\mathrm{R}}}} \left( \frac{m_{\mathrm{N}} E_{\mathrm{R}}}{m_{\chi, \mathrm{N}}} + \delta \right) \, ,
\end{equation}
where $m_{\mathrm{N}}$ is the mass of the target xenon nucleus, and $m_{\chi, \mathrm{N}}$ is the DM-nucleus reduced mass. 
The positive mass splitting $\delta$ kinematically suppresses the interaction rate for low-velocity particles, thereby favoring heavy nuclear targets like xenon and shifting the expected single-scatter nuclear recoil spectrum toward higher energies compared to purely elastic scattering. Additionally, the excited MiDM state is allowed to decay back to the ground state through the emission of a photon of energy $\delta$, with a characteristic lifetime of $\tau=\pi/\delta^3\mu^2_\chi$. This gives rise to a unique signal topology consisting of an initial nuclear recoil followed by a delayed electronic recoil from the photon, depicted in Figure~\ref{fig:topology}, which we explicitly search for in this analysis.

\textit{Detector}---The XENONnT experiment consists of a central cylindrical TPC containing an active target mass of $5.9\,\mathrm{tonnes}$ of liquid xenon, with a radius of $66.4\,\mathrm{cm}$ and a length of $149\,\mathrm{cm}$ \cite{xenonnt2024}. The TPC is located inside nested water Cherenkov neutron \cite{Aprile_2025_NV} and muon \cite{Aprile_2014} vetos, designed to suppress radiogenic neutron backgrounds and cosmogenic muon-induced events. Detector operation is supported by several auxiliary subsystems, including gas- and liquid-phase xenon purification systems \cite{Plante_2022}, a radon distillation system \cite{aprile2025radonremovalxenonntsolar}, and a krypton distillation column \cite{krypton2017}. Interactions within the xenon target give rise to distinct recoil signatures: particles scattering off xenon nuclei produce nuclear recoils (NRs), while interactions with atomic electrons result in electronic recoils (ERs). Scintillation and ionization signals, conventionally denoted as S1 and S2, are detected by arrays of photomultiplier tubes (PMTs) located at the top and bottom of the TPC \cite{Antochi_2021}. PMT signals exceeding the digitization threshold are recorded as individual hits. These hits are subsequently grouped into peaks by dedicated reconstruction algorithms, and their waveforms are converted into units of photoelectrons (PEs) and summed to produce the total waveform of each S1 and S2 peak ~\cite{2023xenonntdaq}.

\textit{Dataset}---This analysis employs the Science Run 1 (SR1) dataset collected with XENONnT, obtained between May 19, 2022, and August 8, 2023, corresponding to a total science exposure of 2.1~$\mathrm{tonne-years}$. The live time of SR1 is subdivided into SR1a (66.6 days) and SR1b (119.9 days), differentiated by different levels of $^{85}$Kr \cite{XENONnT:2025sr1}. SR1 was the longest available science run at the time of this analysis; SR0 was reserved for potential validations and ultimately not included in the search. The radon distillation system was operated in both gaseous xenon (GXe) and liquid xenon (LXe) modes throughout the run. The average electron lifetime, characterizing the survival of drifting electrons against attachment to electronegative impurities, is measured to be $21.8^{+6.7}_{-9.7}\,\mathrm{ms}$, while the detector gains are determined to be $g_1 = (0.1367 \pm 0.0010)\,\mathrm{PE/photon}$ and $g_2 = (16.85 \pm 0.46)\,\mathrm{PE/e^-}$. 

\textit{Signals}---The inelastic nuclear recoil energy spectrum depends on the DM mass $m_\chi$, the mass splitting $\delta$, and the relevant coupling parameter: the spin-independent nucleon cross section $\sigma_{\mathrm{SI}}$ for SIiDM, or the magnetic dipole moment $\mu_\chi$ for MiDM. 
For SIiDM, the differential event rate follows the standard expression:
\begin{equation}\label{eq:siidm_recoil_spectrum}
    \frac{dR}{dE_{\mathrm{R}}}
=
\frac{\rho_0 \,\sigma_{\mathrm{SI}} \, A^2}{2 \, m_\chi \, m^2_{\chi, \mathrm{n}}} 
 F^2(q) 
\int_{v_{\min}(E_{\mathrm{R}},\delta)}^{\infty}
\frac{f(v)}{v}\, d^3 v \, ,
\end{equation}
\noindent where $\rho_0 = 0.3 \text{ GeV/c}^2\text{/cm}^3$ is the local DM density, $A$ is the xenon mass number, $m_{\chi, \mathrm{n}}$ is the DM-nucleon reduced mass, $f(v)$ is the DM velocity distribution (taken from \cite{Baxter_2021}) and $F(q)$ is the Helm nuclear form factor \cite{Lewin1995rx}; the speed of light $c$ is set to 1. For MiDM, the differential rate is instead expressed as a function of the nuclear recoil energy $E_{\mathrm{R}}$, the velocity of the $\chi^*$ DM state after scattering $|\vec{v_*}|$, and $x_*$, defined as the cosine of the angle between $\vec{v_*}$ and the Earth's velocity in the Galactic frame, $\vec{v}_\mathrm{E}$. The full expression for the MiDM differential rate is provided in the End Matter and in the literature \cite{Lin_2011}.
Increasing $\delta$ leads to kinematic suppression of the rate due to larger values of $v_{\min}$, and shifts the spectrum toward higher recoil energies. The overall event rate scales proportionally with $\mu_\chi^2$ in MiDM and with $\sigma_{\mathrm{SI}}$ in SIiDM. Based on the expected event rates, we search for values of $\delta$ between $20\,\mathrm{keV}$ and $250\,\mathrm{keV}$, beyond which the event rate is exceedingly suppressed.

Standard SIiDM does not feature a prompt, observable decay signature within the detector volume~\cite{TuckerSmith:2001hy, Chang:2010en}; therefore, the only relevant signal consists of single-scatter nuclear recoils with inelastic energy spectra. In the case of MiDM, the excited state decays by emitting a photon. If this photon does not deposit energy within the active volume, only the initial nuclear recoil is observed, resulting in a \textit{single-site} interaction analogous to that expected from SIiDM.
Single-site interactions are identified using the same event-building procedure adopted in previous analyses, based on the selection of well-reconstructed S1--S2 pairs. Details of the single-site signal reconstruction are provided in \cite{xenoncollaboration2024xenonntanalysissignalreconstruction}. Nuclear recoil signals in the single-site channel are modeled in the space of the S1 and S2 signals after spatial and time-dependent corrections (cS1 and cS2, respectively) and the reconstructed event radius ($R$), using detector response simulations informed by detector calibration data. The lower energy threshold of this search is intrinsically determined by the 3-fold PMT coincidence requirement for S1 detection \cite{xenoncollaboration2024xenonntanalysissignalreconstruction}. At higher energies, we search for NR signatures up to a $\mathrm{cS1} = 100\,\mathrm{PE}$, corresponding to an nuclear recoil of $[3.8, 64.1]\,\mathrm{keV_{\mathrm{NR}}}$ \cite{XENONnT:2025sr1}, and account for the small signal loss due to this upper threshold in the analysis.

If the MiDM decay takes place inside the active volume the emitted photon will deposit its energy in the form of an ER, resulting in a characteristic \textit{double-site} signal. Figure~\ref{fig:topology} illustrates the distinctive event topology of MiDM: an initial inelastic nuclear recoil (NR) followed by a decay that emits a $\gamma$-ray of energy $\delta$ before the DM particle exits the active volume. This sequence produces both a NR and an ER, yielding two S1 and two S2 signals within a single event window. Given the energy hierarchy ($\delta$ is larger than the NR energy in our ROI), the $\gamma$-induced ER generates the larger S1 and S2 pair. The largest detected S2 is thus labeled as the ER S2, and the second-largest S2 as the NR S2. The NR scintillation signal is searched for in the $5\,\mu\mathrm{s}$ window immediately preceding the ER S1, which corresponds to the maximum time that a DM particle is expected to remain inside of the active volume, based on the dimensions of the detector and the characteristic values of $v_*$. The NR S2 signal can precede or follow the ER S2, as a function of their relative vertical ($z$) locations. In this analysis, we consider both single-site signals (for SIiDM and MiDM) and double-site signals (for MiDM) in order to maximize the scope and sensitivity of the search.

\begin{figure}[!h]
    \centering
    \includegraphics[width=1\linewidth]{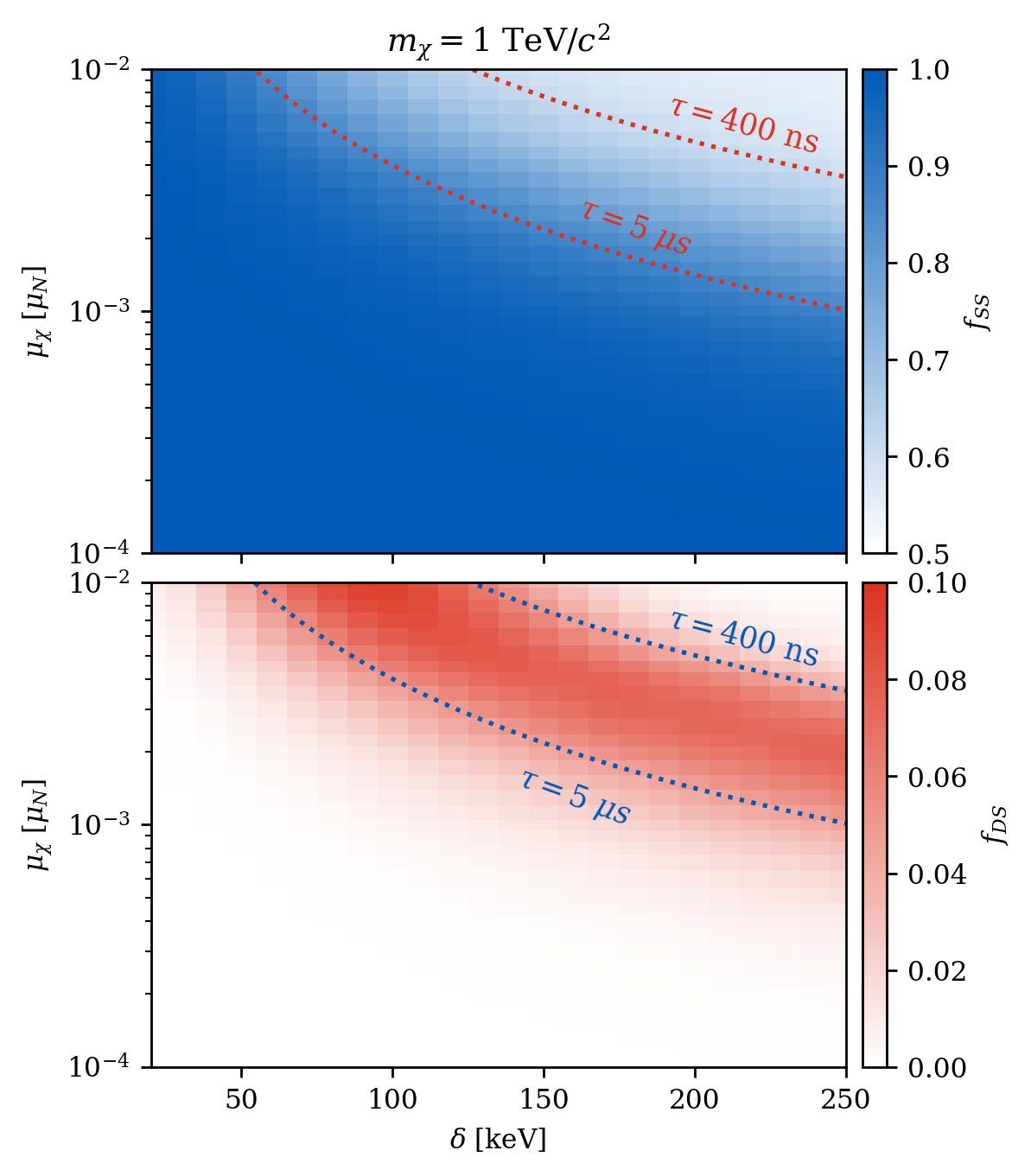}
    \caption{Distribution of MiDM signal fraction between the single-site ($f_{SS}$, top) and double-site ($f_{DS}$, bottom) signal channels, after accounting for detection and selection efficiencies, for a $1\,\mathrm{TeV/c^2}$ particle. Dashed lines indicate contours of constant $\chi^*$ lifetime.}
    \label{fig:signal_fractions}
\end{figure}

\textit{Simulations}---MiDM signals are simulated using the XENONnT full-chain simulation framework FUSE~\cite{fuse} to quantify the fraction of events producing single-site and double-site topologies, and to characterize their properties. These reproduce both the scattering and decay processes of MiDM particles for different values of $m_\chi$, $\delta$, and $\mu_\chi$, recording the time, position, and energy of each interaction. For each event, the decay time is sampled from the characteristic lifetime $\tau$, while $v_*$ is determined from the scattering kinematics.
The fraction of simulated events in which the emitted photon does not interact within the active volume contributes to the single-site channel, denoted $f_{\mathrm{SS}}$. Figure~\ref{fig:signal_fractions} shows the dependence of $f_{\mathrm{SS}}$ on $\delta$ and $\mu_\chi$ for a representative MiDM mass of $m_\chi = 1 \,\mathrm{TeV}/c^2$. As expected, $f_{\mathrm{SS}}$ approaches unity for smaller values of $\delta$ and $\mu_\chi$, reflecting the longer associated $\chi^*$ lifetimes ($\tau \propto \mu_\chi^{-2} \delta^{-3}$). For each interaction, we subsequently simulate the generation of observable quanta and the resulting PMT waveforms, which are processed by the exact same reconstruction algorithm used for real detector data.

For double-site MiDM topologies, the detection efficiency relies not only on the standard data-quality selections for isolated NR and ER signals \cite{xenoncollaboration2024xenonntanalysissignalreconstruction}, but is fundamentally limited by the temporal and spatial resolution of closely spaced signal pairs. To systematically evaluate this reconstruction capability across the target parameter space, we analyze the simulated MiDM waveforms.
The time separation between the two S1 pulses ($\Delta t_{\mathrm{S1}}$) is nearly identical to the decay time of $\chi^*$. Our peak-finding algorithm \cite{xenoncollaboration2024xenonntanalysissignalreconstruction}  successfully resolves distinct S1s with a 50\% probability at $\Delta t_{\mathrm{S1}} = 600\,\mathrm{ns}$, rising to 100\% for delay times approaching $\sim 1.4\,\mu\mathrm{s}$.
Similarly, the merging of S2 pulses depends primarily on their vertical ($z$) proximity. The minimum $z$ separation required to independently reconstruct the NR and ER S2 signals varies between roughly $0.4\,\mathrm{cm}$ and $1.4\,\mathrm{cm}$ depending on the event depth, as determined from detector calibrations. By evaluating the kinematic vertical separation of simulated MiDM events against this depth-dependent resolution, we systematically compute the S2 reconstruction efficiency. For events where the S1 pulses are successfully resolved, we find this S2 efficiency exceeds 96\% across our entire search region. Thus, S2 merging is a subdominant effect compared to S1 merging, though it is fully incorporated into our signal detection efficiency model.
In addition, $\gamma$-rays of energies above $\sim 150\,\mathrm{keV}$ may undergo Compton scatterings, producing fragmented ER S2s. 
Since this morphology is subdominant below our $\delta$ search bound of $250\,\mathrm{keV}$, we bypass specialized multi-site ER reconstruction. Instead, any efficiency loss from such events is intrinsically accounted for via the full-chain simulations; we estimate this selection survival probability to be $>99\%$ for $150\,\mathrm{keV}$ $\gamma$-rays, gradually decreasing to roughly $96\%$ at the $250\,\mathrm{keV}$ upper bound of our search parameter space.
Once the S1 and S2 reconstruction efficiencies are modeled, the fraction of events resulting in properly reconstructed double-site signals, denoted as $f_{\mathrm{DS}}$, is computed; it is shown in Figure~\ref{fig:signal_fractions}. As depicted, only a certain band in ($\delta, \mu_\chi$) yields a considerable population of double-site signals, that with $\tau$ values between $400\,\mathrm{ns}$ and $5\,\mu\mathrm{s}$. Below $400\,\mathrm{ns}$, S1 merging completely suppresses the detection of these events; above  $5\,\mu\mathrm{s}$, the DM particle typically travels far enough that the emitted photon rarely interacts within the XENONnT active volume.

\textit{Event selection}---Single-site events are selected using the established XENONnT SR1 WIMP search criteria \cite{xenoncollaboration2024xenonntanalysissignalreconstruction}. For double-site events, the NR candidate follows identical criteria except for the single-scatter requirement. The ER candidate is subject to standard data-quality cuts, based on the distribution of S1 and S2 area between the top and bottom arrays, S1 and S2 peak widths and S1 PMT contribution to the peak, as well as an S1 single-scatter cut. Both channels utilize the standard SR1 fiducial volume containing $4.05\,\mathrm{tonnes}$ of liquid xenon \cite{XENONnT:2025sr1}. The overall double-site selection acceptance varies between 85\% and 90\%, primarily depending on the ER size.

To suppress backgrounds in the double-site channel, we introduce three novel criteria: the kinematic, velocity, and directional cuts. These exploit the correlated dynamics of genuine MiDM events, discriminating against physical and instrumental backgrounds based on the reconstructed $\chi^*$ speed and direction. 
Crucially, these cuts target two specific background classes: accidental coincidences (ACs) and the delayed coincidences from $^{85}\mathrm{Kr}$ decays, discussed below.  Detailed accounts of the implementation and performance of these selection criteria are provided in the End Matter.

\textit{Backgrounds}---Different background sources are considered for the single-site and double-site signal channels. The single-site background sources and their models in ($\mathrm{cS1}$, $\mathrm{cS2}$, $R$) are identical to those considered for the standard WIMP search in this dataset \cite{XENONnT:2025sr1}; their origin and the methodologies used for modeling them can be found in \cite{Aprile_2025}. Several potential background sources are considered for the double-site channel, including ACs, delayed $^{85}\mathrm{Kr}$ decays, and radiogenic neutron-induced events. ACs can arise through two mechanisms. In the first, an independent ER event overlaps in the same event window with an NR or NR-like event. The NR-like event may be a low-energy ER event that leaks into the NR band. In the second, a lone S1 overlaps with an ER multi-scatter event whose two S2 signals happen to satisfy the NR and ER region-of-interest requirements. Such ER multi-scatter events, e.g.\ from $\gamma$-ray Compton scattering, produce two energy-deposition sites separated by the $\gamma$-ray mean free path in liquid xenon, several times smaller than the separation of an authentic MiDM signal at the relevant energies \cite{Israelashvili_2015}. The resulting apparent velocity therefore lies outside the kinematically allowed MiDM range, causing both AC topologies to fail the kinematic and velocity selections. Delayed coincidences from $^{85}\mathrm{Kr}$ decays occur at the same location, yielding no spatial displacement, and involve energy depositions incompatible with MiDM scattering kinematics. They therefore also consistently fail these criteria and are completely rejected. The AC and $^{85}\mathrm{Kr}$ components are thus rendered negligible.

Radiogenic neutron-induced events constitute the background topology most closely resembling double-site MiDM signals. Unlike the AC and $^{85}\mathrm{Kr}$ backgrounds, its suppression cannot be argued analytically from the cut definitions alone, and its rate must instead be quantified through dedicated simulations. We employ GEANT4 simulations, which begin with the emission of $\alpha$-particles by radioactive contaminants in the detector materials \cite{XENONnT:2022hdt}. These induce a rate of $(\alpha, n)$ processes in the material, resulting in a flux of neutrons towards the liquid xenon active volume. The path and interactions of each emitted neutron is simulated, including recoils, absorptions and subsequent relaxation processes. The S1 and S2 response of all energy depositions in the liquid xenon are then obtained through the same simulation software as the one employed for the simulation of MiDM events. 

The final background rate of neutron-induced double-site events is obtained by applying all MiDM double-site selection cuts to the simulated dataset, and counting the number of surviving events. The estimated rates are $m_\chi$-dependent due to the implementation of the kinematic and velocity cuts, and remain below $0.003 \pm 0.002$ expected background events in SR1, with uncertainties arising from the finite simulations statistics as well as systematic uncertainties in the radio-impurity contents of the detector materials. The double-site MiDM search channel is therefore essentially background-free, such that the detection of a single event would constitute a significant excess above the expectation of the background models.

\begin{figure}[!h]
    \centering
    \includegraphics[width=1\linewidth]{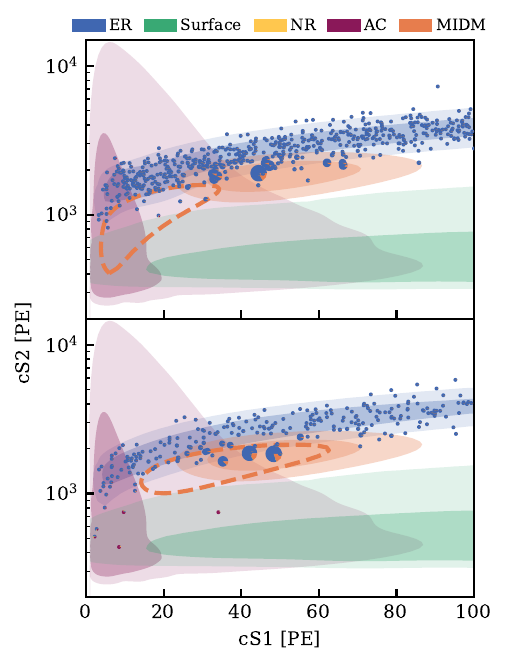}
    \caption{Events remaining after all selection criteria are applied: 805 single-site and 0 double-site events. The top panel depicts events and background templates in the SR1a portion of the dataset, while the bottom panel depicts SR1b. Templates are shown for sources of background (except the NR background, for clarity) as 1 and 2\,$\sigma$ contours, as well as for the best-fit MiDM signal, which corresponds to $m_\chi = 50\,\mathrm{GeV/c^2}$ and $\delta=120\,\mathrm{keV}$. Dashed lines depict the signal template for a MiDM of equal mass and a splitting of $50\,(100)\,\mathrm{keV}$ in the top (bottom) panel. The top (bottom) panel shows events and templates for the SR1a (SR1b) portion of SR1, which differ slightly in the ER background. Each event is represented by a pie chart depicting the likelihood contribution of each event source for the best-fit model.}
    \label{fig:pie_plot}
\end{figure}

\begin{figure}[!h]
    \centering
    \includegraphics[width=0.98\linewidth]{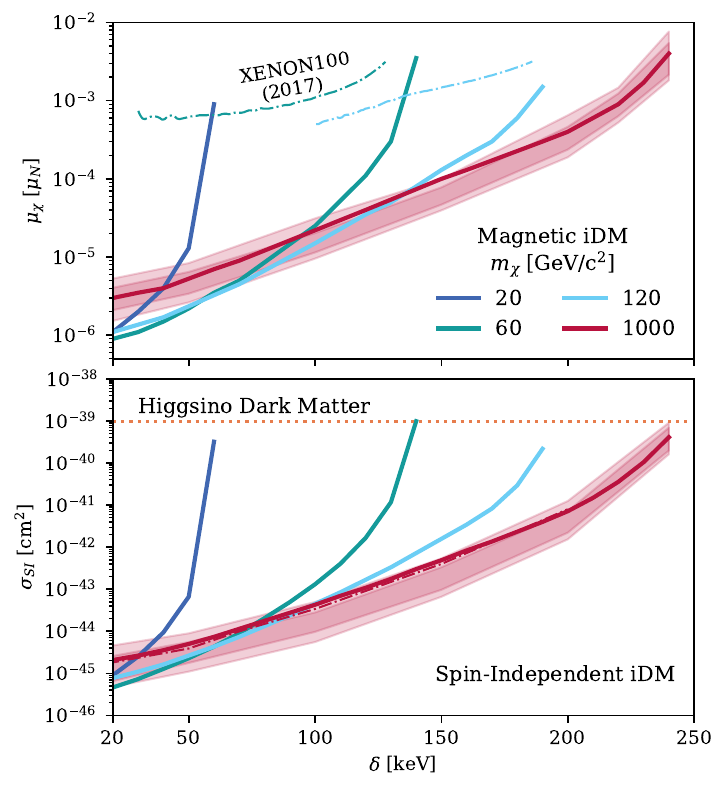}
    \caption{Upper limits at 90\% C.L. on the values of $\mu_\chi$ for MiDM (top) and $\sigma_{\mathrm{SI}}$ for SIiDM (bottom), as a function of $\delta$, for different $m_\chi$. The corresponding sensitivity bands for $m_\chi = 1 \text{ TeV/c}^2$ is indicated by the region containing 68 \% and 95\% 
    of expected upper limits under the background-only hypothesis. Previous best limits are shown (dashed) for both models from the XENON100 \cite{Aprile_2017} and XENON1T~\cite{Aprile_2024} experiments, respectively. The spin-independent cross section for Higgsino DM (\cite{PhysRevD.111.055030}) is shown in orange.}
    \label{fig:limits}
\end{figure}

\textit{Inference}---We develop a combined likelihood function in order to combine the single-site and the double-site search channels at the statistical level in the MiDM search. Because the single-site and double-site search channels are mutually exclusive, their respective likelihood functions are multiplied,

\begin{equation}
    \mathcal{L}(\mu_\chi, \bm{\theta}) = \mathcal{L}_{\mathrm{SS}}(\mu_\chi, \bm{\theta}) \times \mathcal{L}_{\mathrm{DS}}(S(\mu_\chi) + B(\bm{\theta}), \bm{\theta}) \, ,
\end{equation}

\noindent where $\bm{\theta}$ denotes a set of ancillary parameters used to fit the background and signal response model to the data. The single-site likelihood term $\mathcal{L}_{\mathrm{SS}}$ is identical to the one presented in the search for the standard WIMP search; it depends on the signal and background models in ($\mathrm{cS1}$, $\mathrm{cS2}$, $R$) space, and makes use of $^{220}\mathrm{Rn}$ calibration data to fit the shape of the ER band; details about the statistical approach in this channel can be found in \cite{Aprile_2025}.

Given the low backgrounds expected in the double-site search channel, we simplify the double-site likelihood term $\mathcal{L}_{\mathrm{DS}}$ to a Poisson function $\mathrm{Pois}(N_{\mathrm{DS}} | \mu^{\mathrm{DS}}_{\mathrm{tot}}(\mu_\chi, \bm{\theta}))$, where $N_{\mathrm{DS}}$ is the number of events observed in the double-site channel and $\mu^{\mathrm{DS}}_{\mathrm{tot}}$ is the number of expected events. Because SIiDM does not feature a decay signature, we drop this term for this search and employ only the single-site likelihood term.

Sensitivity studies based on simulations indicate that the single-site search channel dominates for value of $\delta < 180$ keV, while the double-site search channel dominates for values above that. The combined likelihood thus maximizes the sensitivity across all values of $\delta$ in the ROI. 

\textit{Results}--- Double-site events were kept blinded during the development of selection criteria and the modeling of background. After unblinding and applying all selection cuts, 0 double-site events were observed in SR1. Single-site selection criteria were left unchanged from~\cite{XENONnT:2025sr1}, which were also developed blindly. Figure~\ref{fig:pie_plot} depicts the 805 single-site events, together with the relevant signal and background templates.

We find no statistically significant indication of an iDM-like excess in our dataset. The best-fit models of MiDM and SIiDM to the data both correspond to $m_\chi = 50\, \mathrm{GeV/c^2}$ and $\delta = 120\, \mathrm{keV}$, with $\mu_\chi = 2.5 \times 10^{-4} \, \mu_\mathrm{N}$ and $\sigma_{SI} = 1.5 \times 10^{-41}\,\mathrm{cm^2}$, respectively (where $\mu_\mathrm{N}$ is the nuclear magneton); both have a statistical significance of $1.6\,\sigma$ over the background-only fit. The similarity in the best-fit properties of the MiDM and SIiDM models is due to the lack of double-site events in our dataset, and the similar nuclear recoil spectra between the two. We place upper limits on the values of $\mu_\chi$ and $\sigma_{\mathrm{SI}}$ for different values of $m_\chi$ and $\delta$. Figure~\ref{fig:limits} shows the resulting 90\% C.I. upper bounds for a selection of values of $m_\chi$. We improve on current limits \cite{Aprile_2017} of the iDM magnetic dipole moments by up to 3 orders of magnitude, excluding values above $9\times 10^{-7}\, \mu_\mathrm{N}$ for a $50\,\mathrm{GeV/c^2}$ mass and $20\,\mathrm{keV}$ mass splitting.

\textit{Conclusion}---In this Letter, we have presented a search for Magnetic and Spin-Independent inelastic Dark Matter using $2.1\,\mathrm{tonne-years}$ of science exposure in the XENONnT detector. We find no indication of an iDM-like excess in our dataset, which allows us to place world-leading upper limits on the values of the spin-independent DM-nucleon cross section and on the value of the Dark Matter magnetic dipole moment. 

\vspace{1em}
\begin{acknowledgments}

We gratefully acknowledge support from the National Science Foundation, Swiss National Science Foundation, German Ministry for Education and Research, Max Planck Gesellschaft, Deutsche Forschungsgemeinschaft, Helmholtz Association, Dutch Research Council (NWO), Fundacao para a Ciencia e Tecnologia, Weizmann Institute of Science, Binational Science Foundation, Région des Pays de la Loire, Knut and Alice Wallenberg Foundation, Kavli Foundation, JSPS Kakenhi, JST FOREST Program, and ERAN in Japan, Tsinghua University Initiative Scientific Research Program, National Natural Science Foundation of China, Ministry of Education of China, DIM-ACAV+ Région Ile-de-France, and Istituto Nazionale di Fisica Nucleare. This project has received funding/support from the European Union’s Horizon 2020 and Horizon Europe research and innovation programs under the Marie Skłodowska-Curie grant agreements No 860881-HIDDeN and No 101081465-AUFRANDE.
We gratefully acknowledge support for providing computing and data-processing resources of the Open Science Pool and the European Grid Initiative, at the following computing centers: the CNRS/IN2P3 (Lyon - France), the Dutch national e-infrastructure with the support of SURF Cooperative, the Nikhef Data-Processing Facility (Amsterdam - Netherlands), the INFN-CNAF (Bologna - Italy), the San Diego Supercomputer Center (San Diego - USA) and the Enrico Fermi Institute (Chicago - USA). We acknowledge the support of the Research Computing Center (RCC) at The University of Chicago for providing computing resources for data analysis.
We thank the INFN Laboratori Nazionali del Gran Sasso for hosting and supporting the XENON project.

\end{acknowledgments}

\bibliography{references}

\appendix

\section{End Matter}

\subsection{MiDM Differential Rate}\label{app:midm_rate}
The three-dimensional differential rate of MiDM recoils as a function of $E_{\mathrm{R}}$, $v_*$ and $x_*$ is given by the expression

\begin{equation}
\begin{aligned}
\frac{d^3 R}{d E_{\mathrm{R}} d v_* d x_*} &= \frac{\eta N_{\mathrm{T}} \rho_\chi \hbar^2 }{m_{\mathrm{N}}} \frac{|\mathcal{M}|^2}{32 \pi m_\chi^3} F^2(E_{\mathrm{R}}) \\
&\hfill \left[ v_* \Theta(1-|x_q|) \int d \phi \frac{e^{-(v^{\prime})^2 / v_0^2}}{n(v_0, v_{\mathrm{esc}})} \Theta(v_{\mathrm{esc}}-|\vec{v}^{\prime}|) \right],
\end{aligned}
\end{equation}
\noindent where $x_q$ corresponds to the cosine of the angle between the direction of the recoiling nucleus and $v_*$, $\vec{v'}=\vec{v}+\vec{v}_{\mathrm{E}}$ corresponds to the velocity of the incoming DM particle in the galactic frame and $\phi$ is the angle between $\vec{v}$ and $\vec{v}_{\mathrm{E}}$:
\begin{equation}
x_q = -\frac{E_{\mathrm{R}} \left(\frac{m_{\mathrm{N}}}{m_\chi} - 1\right) - \delta}{q v_*}, \quad \text{and}
\end{equation}

\begin{equation}
\begin{aligned}
(v')^2 &= v_{\scriptscriptstyle \mathrm{E}}^2
       + \frac{q^2}{m_\chi^2}
       + v_*^2
       + 2 v_{\scriptscriptstyle \mathrm{E}} v_* x_* 
       + \frac{2 x_q q v_*}{m_\chi} \\
&\hfill+ \frac{2 v_{\scriptscriptstyle \mathrm{E}} q}{m_\chi}
\left(
x_q x_*
+ \sqrt{1-x_q^2}\sqrt{1-x_*^2}\cos\phi
\right).
\end{aligned}
\end{equation}

A complete derivation of the expressions above can be found in \cite{Lin_2011}, including expressions for the matrix elements $|\mathcal{M}|^2$, which encode the dependence of the event rate with $\mu_\chi^2$.

\subsection{MiDM Double-Site Selection Criteria}\label{app:novel_cuts}

\begin{figure}[!h]
    \centering
    \includegraphics[width=1\linewidth]{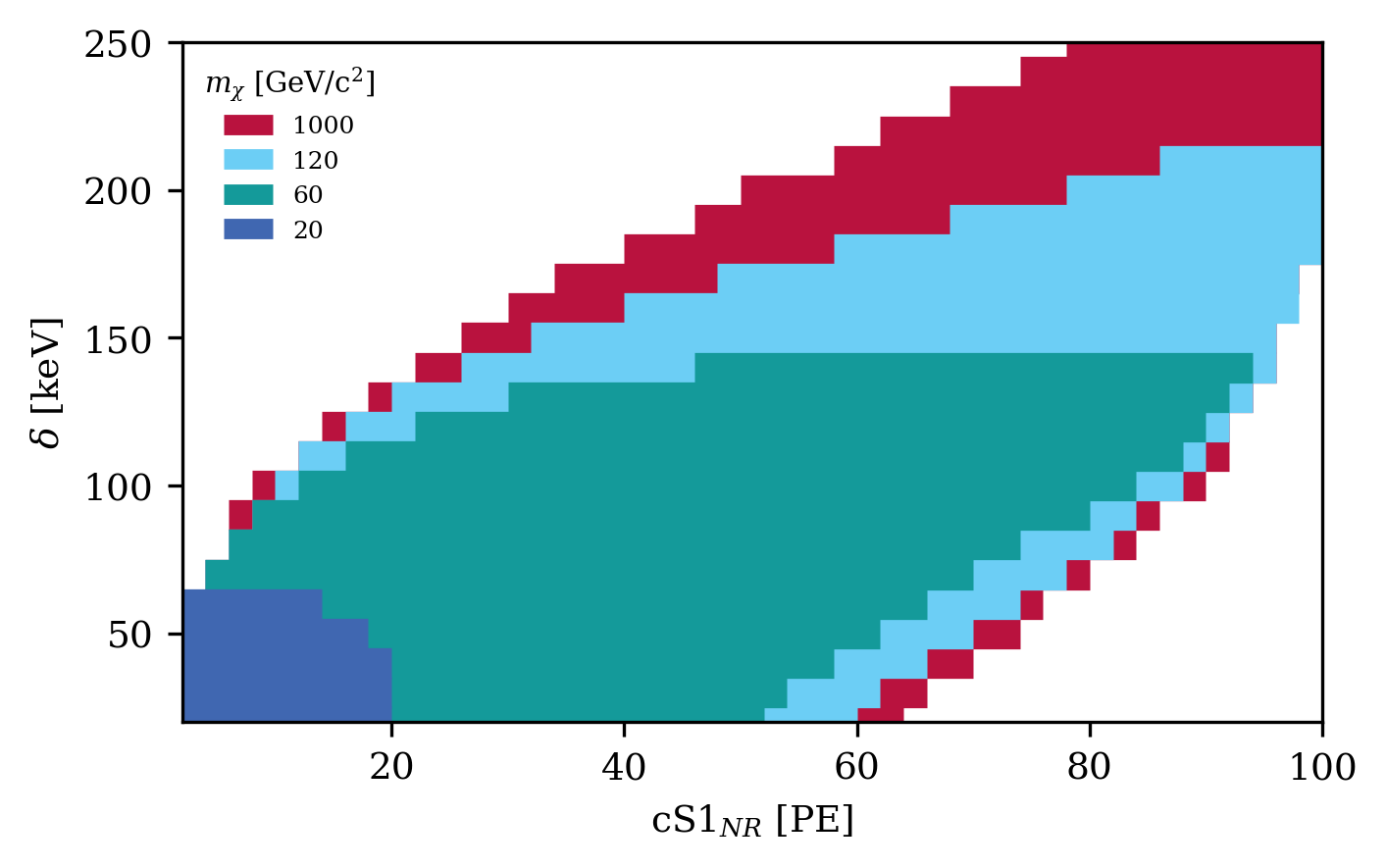}
    \caption{Boundaries of the kinematic selection criteria for a 95\% signal acceptance and a selection of DM masses.}
    \label{fig:kinematic}
\end{figure}

\noindent \textit{Kinematic cut}---The kinematic selection exploits the expected correlation between $\delta$ and the nuclear recoil energy $E_{\mathrm{R}}$. As $\delta$ increases, the scattering process becomes more endothermic, requiring a larger momentum transfer between the DM particle and the xenon nucleus to conserve energy and momentum. This results in a positive correlation between $\delta$ and $E_{\mathrm{NR}}$, a condition that background events are not required to satisfy.
The value of $\delta$ can be inferred from the ER candidate, while $E_{\mathrm{NR}}$ corresponds to the reconstructed recoil energy of the NR candidate. Consequently, the compatibility between the NR and ER components in a double-site event can be assessed based on their respective energies.
Using the ($\mathrm{cS1}$, $\mathrm{cS2}$, $R$) templates for MiDM nuclear recoils developed for the single-site signal search, we determine the central 95\% $\mathrm{cS1}_{\mathrm{NR}}$ interval as a function of $m_\chi$ and $\delta$, and select only events within this interval. This procedure yields a signal acceptance of 95\%, with an estimated background rejection efficiency of 20--25\%, depending on $m_\chi$. Figure~\ref{fig:kinematic} depicts the boundaries of the kinematic selection criteria, as a function of the MiDM mass, reconstructed mass splitting and the size of the S1$_{\text{NR}}$.

\noindent \textit{Velocity cut}---In a MiDM event, the excited state $\chi^*$ propagates along a straight trajectory between the scattering vertex and the decay site. The magnitude of its velocity $v_*$ is determined by the DM halo velocity distribution \cite{Baxter_2021} and the kinematics of the scattering process. 
The velocity selection criterion identifies candidate events based on the apparent velocity of an NR--ER pair, where $v_*$ is reconstructed from the measured positions $\vec{r}_i$ and times $t_t$ of the NR and ER interaction sites:
\begin{equation}
    \hat{v}_* =
    \frac{|\Delta \vec{r}|}
         {\Delta t_{\scriptscriptstyle \mathrm{S1}}}
    =
    \frac{|\vec{r}_{\scriptscriptstyle \mathrm{ER}}
          - \vec{r}_{\scriptscriptstyle \mathit{NR}}|}
         {t_{\scriptscriptstyle \mathrm{S1,ER}}
          - t_{\scriptscriptstyle \mathrm{S1,NR}}}
\end{equation}
For each candidate NR -- ER pair, the velocity of $\chi^*$ can be calculated and compared to the kinematically allowed values of $v_*$, which are given by 
\begin{equation}\label{eq:vstarmax}
    v_{*}^{\max}(E_{\mathrm{R}}) = \sqrt{(v_{\mathrm{E}} + v_{\mathrm{esc}})^2 - 2(E_{\mathrm{R}} + \delta)/m_\chi},
\end{equation}
and 
\begin{equation}\label{eq:midm_min_vel}
    v_*^{\min}(E_{\mathrm{R}}) = \Bigg |\frac{E_{\mathrm{R}}(\frac{m_{\mathrm{N}}}{m_\chi}-1)-\delta}{\sqrt{2m_{\mathrm{N}}E_{\mathrm{R}}}}\Bigg|
\end{equation}
Events with apparent velocities outside of the kinematically allowed bounds are deemed unphysical and excluded from the dataset. In addition, we impose the condition $v_* \geq 100\,\mathrm{km/s}$, which removes approximately 50\% of the neutron-induced background events according to our simulated datasets. This induces a small loss of MiDM acceptance (below 1\%) which we account for in the analysis. The reconstruction accuracy of $v_*$ in MiDM events is corroborated through simulations, for which the true value of $v_*$ is known. We compare the reconstructed value of $v_*$ to its true value, finding very good agreement between the two with negligible errors. 

\begin{figure}[!h]
    \centering
    \includegraphics[width=1\linewidth]{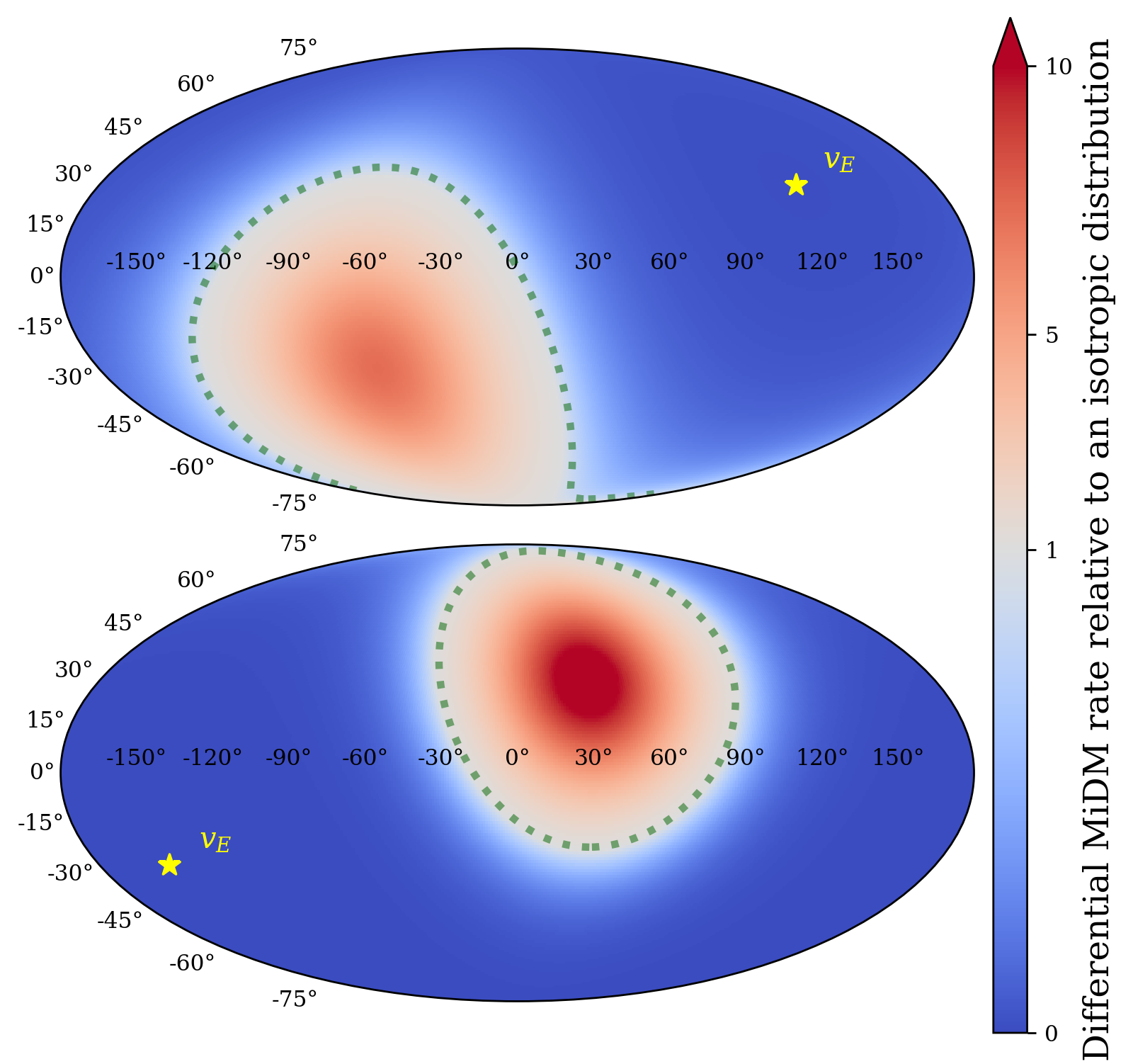}
    \caption{Distribution of MiDM directional signal density over the sky sphere for $v_* = 200$ km/s (top) and $v_* = 500$ km/s (bottom), as a multiple of the signal density of an isotropic distribution. The placement of the directional cut is shown in dashed green lines, for two different values of $v_*$.}
    \label{fig:directional}
\end{figure}

\noindent \textit{Directional cut}---The directional selection exploits the Earth's motion in the Galactic frame ($\vec{v}_{\mathrm{E}}$) to discriminate MiDM signal events from background. This motion induces a \textit{DM wind}, such that DM particles are expected to traverse the detector preferentially in the direction opposite to $\vec{v}_{\mathrm{E}}$. The vector connecting the NR and ER interaction sites of a candidate event can therefore be used to discriminate signal from background, based on whether the direction of the event matches the expectations.
The resulting directional clustering induced by $\vec{v}_{\mathrm{E}}$ depends on the magnitude of $v_*$, becoming more concentrated with higher $|v_*|$ values. Figure~\ref{fig:directional} illustrates the expected angular distribution of MiDM signal events over the sky sphere, for two representative orientations of $\vec{v}_{\mathrm{E}}$ and for two values of $|v_*|$.

The location of $\vec{v}_{\mathrm{E}}$ on the sky sphere is computed for the time of each candidate delayed-coincidence event, which yields the DM signal distribution for the measured $|v_*|$ of the event. The directional cut is then applied based on the NR$\rightarrow$ER vector direction, accepting only the fraction of the sky in which the event density is higher than that of an isotropic distribution, optimizing between signal acceptance and background rejection. 

The neutron-induced background of the double-site channel is assumed to be isotropically distributed in the galactic frame, given that the detector rotates in this frame once every day. Thus, over the duration of SR1, any directionality of the neutron background in the detector's frame becomes isotropic when measured in the galactic frame. As a result, the directionality selection cut achieves a signal acceptance of 86-93\% and a background rejection efficiency of 70-85\%, as a function of the value of $m_\chi$ and $\delta$.

Similarly to the velocity cut, the accuracy with which the path of the $\chi^*$ particle can be reconstructed by the detector is evaluated using simulations. Here, effects like the finite propagation of the photon emitted at decay could yield errors in reconstruction. We find that the median angular error decreases with the length of the distance transversed by $\chi^*$, ranging from errors of $4^{+5}_{-3}\,\mathrm{degrees}$ for path lengths of 20 cm to $1^{+2}_{-1}\,\mathrm{degrees}$ for path lengths of $150\,\mathrm{cm}$. While small, these reconstruction errors are accounted for in the implementation of the directionality cut.

\end{document}